\documentclass[12pt]{article}
\pdfoutput=1
\usepackage{cite}
\usepackage{amssymb}
\usepackage{amsmath}
\usepackage{bm} 
\usepackage{graphicx}
\usepackage{color}
\usepackage{xcolor}
\usepackage{dsfont}
\usepackage{enumerate}
\usepackage{hyperref}
\usepackage{setspace}
\usepackage[latin1]{inputenc}
\definecolor{nicered}{rgb}{0.7,0.1,0.1}
\definecolor{nicegreen}{rgb}{0.1,0.5,0.1}
\hypersetup{colorlinks,citecolor= nicegreen,linkcolor= nicered}
\nopagebreak
\allowdisplaybreaks
\begin{document}
\begin{titlepage}
  \newcommand{\AddrIPM}{{\sl \small School of physics, Institute for
      Research in Fundamental Sciences (IPM),\\ \sl \small
      P.O. Box 19395-5531, Tehran, Iran}}
  \newcommand{\AddrNYU}{{\sl \small New York University Abu Dhabi,\\ \sl \small
      P.O. Box 129188, Saadiyat Island, Abu Dhabi, United Arab Emirates}}
  \vspace*{0.5cm}
\begin{center}
  \textbf{\large Neutrino Non-standard Interactions\\with arbitrary couplings to $\boldsymbol{u}$ and $\boldsymbol{d}$ quarks}\\
  \vspace*{0.4cm}
    Nicol\'as Bernal\footnote{\href{mailto:nicolas.bernal@nyu.edu}{\tt nicolas.bernal@nyu.edu}}
  \vspace*{0.2cm}\\
  \AddrNYU.
  \vspace{0.8cm}\\
    Yasaman Farzan\footnote{\href{mailto:yasaman@theory.ipm.ac.ir}{\tt yasaman@theory.ipm.ac.ir}}
  \vspace*{0.2cm}\\
  \AddrIPM.
  \vspace{1cm}\\
\end{center}
\vspace*{0.2cm}
\begin{abstract}
We introduce a model for Non-Standard neutral current Interaction (NSI) between neutrinos and the matter fields, with an arbitrary coupling to the up and down quarks. The model is based on a new $U(1)$ gauge symmetry with a light gauge boson that mixes with the photon. We show that the couplings to the $u$ and $d$ quarks can have a ratio such that the contribution from NSI to the Coherent Elastic Neutrino-Nucleus Scattering (CE$\nu$NS) amplitude vanishes, relaxing the bound on the NSI from the CE$\nu$NS experiments. 
Additionally, the deviation of the measured value of the anomalous magnetic dipole moment of the muon from the standard-model prediction can be fitted.
The most limiting constraints on our model come from the search for the decay of the new gauge boson to $e^-e^+$ and invisible particles, carried out by NA48/2 and NA64, respectively. We show that these bounds can be relaxed by opening up the decay of the new gauge boson to new light scalars that eventually decay into the $e^- e^+$ pairs. We show that there are ranges that can lead to both a solution to the $(g - 2)_\mu$ anomaly and values of $\epsilon_{\mu \mu} = \epsilon_{\tau \tau}$ large enough to be probed by future solar neutrino experiments.

\end{abstract}
\end{titlepage}
\setcounter{footnote}{0}
\section{Introduction}

Within the Standard Model (SM) of the elementary particles, apart from gravity, the only interaction that neutrinos have is through the weak coupling. With the ever-increasing sensitivity of neutrino experiments, it is timely to ask whether there are any new subdominant interactions between neutrinos and matter fields. In recent years, a remarkable number of studies has been carried on on the impact of neutral current Non-Standard Interaction (NSI) on neutrino propagation in matter. The neutral current NSI can be parameterized as a four-fermion interaction
\begin{equation} 
    2\sqrt{2}\, G_F\, \varepsilon^f_{\alpha \beta} \left(\bar{\nu}_\alpha\, \gamma^\mu\, \frac{1-\gamma_5}{2}\, \nu_\beta\right) \left(\bar{f}\, \gamma_\mu\, (1 + \kappa\, \gamma^5)\, f\right), \label{NSI}
\end{equation} 
where $f\in \{u,d,e\}$. $\varepsilon^f_{\alpha \beta} $ are dimensionless parameters that quantify the strength of the NSI, and the limit $\varepsilon_{\alpha \beta}^f = 0$ corresponds to the standard coupling. In the case where $|\varepsilon^f_{\alpha \beta} |\sim 1$, NSI becomes as strong as the weak interaction.
It is straightforward to show that the axial part of NSI (i.e. the one proportional to $\kappa$) cannot induce matter effects for the propagation of neutrinos in an unpolarized medium such as Earth or the Sun. Moreover, the Coherent Elastic Neutrino-Nucleus Scattering (CE$\nu$NS) experiments are mainly sensitive to the vector part of the NSI. However, the measurement of total solar neutrino flux by SNO was sensitive only to the axial NSI with the quarks. That is, the SNO measurement of the Gamow-Teller process $\nu+D\to \nu+n+p$ can constrain the products $\kappa\times \varepsilon_{\alpha \beta}^u$ and $\kappa\times \varepsilon_{\alpha \beta}^d$ rather than $\varepsilon_{\alpha \beta}^u$ and $\varepsilon_{\alpha \beta}^d$. Measurements of solar neutrino scattering off electrons can constrain both $\varepsilon_{\alpha \beta}^e$ and $\kappa\, \varepsilon_{\alpha \beta}^e$ by studying the dependence of the scattering cross section of the electron recoil energy. 
In this paper, we focus on model building for vector-like NSI, so we fix $\kappa = 0$. 

The effective Lagrangian shown in Eq.~\eqref{NSI} can be obtained by integrating out a heavy $U_\text{NEW}(1)$ gauge boson that couples both to neutrinos and to matter fields. This idea has been pursued in several studies; see, e.g., Ref.~\cite{Farzan:2015doa}. Concerning the propagation of neutrinos in matter, only forward scattering with vanishing energy-momentum transfer ($q^2 \to 0$) is relevant so that here again one can integrate out the mediator and use the effective action in Eq.~\eqref{NSI}, even if the energy of the neutrino beam in the rest frame of the medium is much larger than the mediator mass.\footnote{Indeed, as long as the mass of the mediator is larger than the inverse of the size of the medium, we can integrate out the mediator and rely on  the four-Fermi effective potential formalism~\cite{Smirnov:2019cae}.}

In the presence of NSI, new degeneracies appear in the neutrino oscillation parameters. For example, the so-called generalized mass ordering degeneracy appears~\cite{Gonzalez-Garcia:2013usa, Agarwalla:2014bsa, Bakhti:2014pva, Coloma:2016gei, Liao:2016orc} which leads to an alternative solution to the solar neutrino anomaly known as the LMA-Dark solution with $\theta_{12}>45^\circ$ and $\varepsilon_{\mu \mu}^f-\varepsilon_{ee}^f\simeq \varepsilon_{\tau \tau}^f-\varepsilon_{ee}^f\sim 1$.
As pointed out in Ref.~\cite{Farzan:2017xzy}, if we want to test the LMA-Dark solution via only oscillation experiments, different media with different proton-to-neutron compositions are required. Furthermore, NSI with $|\varepsilon^f| \gtrsim 0.1$ can be tested in principle in scattering experiments.
There are, however, a few exceptions: $i)$ In scattering experiments, if the mediator mass $m_{Z'}$ is smaller than the typical energy-momentum transfer ($\sqrt{|q^2|}$), we cannot use the four-Fermi analysis and we should employ the whole propagator of the mediator that gives an amplitude proportional to $g_{Z'}^2/(m_{Z'}^2-q^2)$ rather than to $g_{Z'}^2/m_{Z'}^2$, and hence a suppression of $m_{Z'}^2/(m_{Z'}^2-q^2)$. $ii)$ With a given target at CE$\nu$NS experiments, the contributions of NSI to the amplitudes of the scattering off the neutrons and protons of the target cancel out each other for certain ratios of $\varepsilon^u_{\alpha \beta}/\varepsilon^d_{\alpha \beta}$~\cite{Esteban:2018ppq, Chaves:2021pey}. Motivated by this phenomenological consideration, we build a model for NSI with an arbitrary ratio of NSI couplings to the $u$- and $d$-quarks.  The scenario is based on a flavor gauge model with a light gauge boson, $Z'$, which mixes with the photon.
We enumerate the relevant bounds on the parameters of the model. 

We focus on the allowed range of the parameter space that can $i)$ explain the $(g-2)_\mu$ anomaly~\cite{Aoyama:2020ynm, Muong-2:2021ojo}, $ii)$ lead to large NSI, and $iii)$ yield ratios of $\varepsilon^u/\varepsilon^d$ for which CE$\nu$NS bounds can be relaxed~\cite{Chaves:2021pey}. In our model, as in the case of $B-L$, the new gauge boson couples to electrons and neutrinos, so it can appear in the NA64 experiment as a missing energy on which there are strong bounds~\cite{Banerjee:2019pds, Andreev:2022hxz}. We discuss how the model can be augmented to suppress the invisible decay modes of $Z'$ and, therefore, open the parameter space to accommodate the solution to $(g-2)_\mu$ and a large NSI.

The paper is organized as follows. In Sect.~\ref{model}, the model is presented. It is also shown how to augment the model to suppress the branching ratios of $Z'\to e^-e^+$ and $Z'\to {\rm invisible}$ in order to avoid the bounds from searches for these decay modes. In Sect.~\ref{bound}, various observables that can test the model are discussed, and the relevant bounds are reviewed. Figures displaying the bounds on the parameter space of our model are presented. The results are summarized in Sect.~\ref{DIS}.

\section{The model \label{model}}
We will augment the SM gauge group with a new local $U_\text{NEW}(1)$ to obtain NSI. We show the lepton and baryon numbers of the three generations with $L_\alpha$ and $B_i$, respectively.
For any arbitrary real value of $c$, the combination of lepton and baryon numbers
\begin{equation}\label{lep-Bar}
    L_\mu+L_\tau -c\, (B_1+B_2) -2\, B_3\, (1-c)
\end{equation} 
is anomaly-free.
The gauge boson of the $U_\text{NEW}(1)$ symmetry is denoted by $Z^\prime$, with a gauge coupling $g_{Z'}$. 
Unless $c=2/3$, the $U_\text{NEW}(1)$ charges of the third generation of quarks are different from those of the first and second generations. As a result, on the quark mass basis, left-handed down quarks can obtain a flavor-violating coupling to $Z'$. This feature has been invoked in Ref.~\cite{Crivellin:2015lwa} to address the so-called $b$-anomalies observed at the LHCb.\footnote{Very recently, the LHCb collaboration reported measurements of the lepton flavor universality in $b \to s\, l^+ l^-$, which for many years dominated the $B$-physics anomalies, compatible with the SM prediction~\cite{LHCb:2022qnv, LHCb:2022zom}.}
We shall comment on whether, in the range of parameters of our interest, the deviation of $b \to s\, \mu^+\, \mu^-$ from the SM prediction is within the observed range or not. We have taken equal charges for the first and second generations of the quarks to respect the bounds from the neutral-kaon mixing.
In the lepton sector, the charged lepton mass basis and the electroweak basis coincide, so we shall not have lepton flavor violating coupling for the charged leptons, but we can have off-diagonal couplings in the neutrino mass basis like $g_{Z'}\, (\delta_{ij}-U_{ei}\, U_{ej}^*)\, Z'_\mu\, \bar{\nu}_i\, \gamma^\mu\, \nu_j$, where $U$ is the PMNS mixing matrix.
This can lead to three-body decay of the heavy neutrinos to lighter ones, but with lifetimes much larger than the age of the universe, an effect irrelevant for phenomenological purposes. Notice that we have set the new $U_\text{NEW}(1)$ charge of the first generation of leptons equal to zero. As a result, the strong limits of GEMMA on $\bar{\nu}_e+e$ scattering can be relaxed~\cite{GEMMA, Harnik:2012ni}. 
The gauge symmetry in Eq.~\eqref{lep-Bar} induces equal couplings to the $u$ and $d$ quarks. We break this universality by introducing a kinetic mixing between $Z'$ and the photon parameterized by $\epsilon$.
The couplings of quarks to $Z'$ can then be written as 
\begin{equation} 
    \left[\left(-\frac{c}{3}\, g_{Z'}+\frac{2}{3}\, e\, \epsilon\right) \sum_{u_i\in \{u,c\}}\bar{u}_i\, \gamma^\mu\, u_i + \left(-\frac{c}{3}\, g_{Z'}-\frac{1}{3}\, e\, \epsilon\right) \sum_{d_i\in \{d,s\}}\bar{d}_i\, \gamma^\mu\, d_i\right] Z'_\mu\,,
\end{equation} 
and the couplings of leptons as
\begin{equation} 
    \left[\left(g_{Z'}-e\, \epsilon\right)\bar\mu\, \gamma^\mu\, \mu + g_{Z'}\, \bar{\nu}_\mu\, \gamma^\mu\, \nu_\mu + \left(g_{Z'}-e\,\epsilon\right) \bar\tau\, \gamma^\mu\, \tau + g_{Z'}\, \bar{\nu}_\tau\, \gamma^\mu\, \nu_\tau - e\, \epsilon\, \bar{\mathbf{e}}\, \gamma^\mu\,   \mathbf{e}\right] Z'_\mu  \ ,
\end{equation} 
where $e$ and $\mathbf{e}$ respectively denote the electric charge and the electron field.
As shown in the appendices of Ref.~\cite{Feldman:2007wj}, as long as there is no mass mixing between $Z'$ and the hypercharge boson, the kinetic mixing cannot induce an electric charge for neutrinos, on which there are extremely strong bounds~\cite{Cadeddu:2019hvy}. Furthermore, in the absence of mass mixing in the St\"uckelberg mass term for the new gauge boson, the bounds from violation of atomic parity do not constrain $\epsilon$~\cite{Davoudiasl:2012ag}. Ref.~\cite{Greljo:2022dwn} has also invoked a $L_\mu-L_\tau$ model with a gauge boson kinetically mixed with the photon that can explain the $(g-2)_\mu$ anomaly.

Integrating out the $Z'$ boson, we can write the following effective couplings to quarks
\begin{align}
	&{\varepsilon}_{\mu\mu}^u = 	\varepsilon_{\tau\tau}^u= \frac{(2\, e\, \epsilon -c\, g_{Z'})\, g_{Z'}}{6\sqrt{2}\, G_F\, m_{Z'}^2}\,,  \label{epU}\\
	&\varepsilon_{\mu\mu}^d = 	\varepsilon_{\tau\tau}^d=- \frac{(e\, \epsilon +c\, g_{Z'})\, g_{Z'}}{6\sqrt{2}\, G_F\, m_{Z'}^2}\,, \label{epD}\\
	&\varepsilon_{ee}^u =	\varepsilon_{ee}^d=0\,,
\end{align}
and to electrons
\begin{align}
    \varepsilon_{ee}^e &= 0\,,\\
    \varepsilon_{\mu \mu}^e &= \varepsilon_{\tau \tau}^e =-\frac{g_{Z'}\, e\,  \epsilon}{2\sqrt{2}\, G_F\, m_{Z'}^2}\,.
    \label{eee}
\end{align}
From Eqs.~\eqref{epU} and~\eqref{epD}, one can obtain the effective couplings to neutrons and protons
\begin{equation} 
    \varepsilon_{\mu \mu}^n=\frac{-c\, g_{Z'}^2}{2\sqrt{2}\, G_F\, m_{Z'}^2}\,,  \  \  \  {\rm and } \ \ \  \varepsilon_{\mu \mu}^p=\frac{(e\, \epsilon-c\, g_{Z'})\, g_{Z'}}{2\sqrt{2}G_F m_{Z'}^2}\  ,
\end{equation} 
and their ratio $\tan \eta$~\cite{Chaves:2021pey}
\begin{equation} 
    \tan \eta=\frac{\varepsilon_{\mu \mu}^n}{\varepsilon_{\mu \mu}^p}=\frac{-c\, g_{Z'}}{e\, \epsilon -c\, g_{Z'}}\,.
\end{equation} 

In this model, the contribution from NSI to the effective potential of neutrinos in matter takes the form $V^\text{NSI}={\rm Diag}(0, V_\mu, V_\tau)$, with
\begin{equation} 
    V_\mu=V_\tau=2\sqrt{2}\, G_F\, (N_e\, \varepsilon_{\mu \mu}^e + N_u\, \varepsilon_{\mu \mu}^u + N_d\, \varepsilon_{\mu \mu}^d)= 2\sqrt{2}\, G_F\, N_e\, \varepsilon_{\mu \mu}^\text{medium}\,,
\end{equation} 
in which
\begin{equation} 
    \varepsilon_{\mu \mu}^\text{medium}=\frac{-c\, g_{Z'}^2}{2\sqrt{2}\, G_F\, m_{Z'}^2}\, \frac{N_n+N_p}{N_e}\ . \label{medium}
\end{equation} 
Notice that we have used the fact that the medium is electrically neutral, so that
\begin{equation} 
    \frac{2}{3}N_u-\frac{1}{3}N_d-N_e=0\  .
\end{equation} 
Taking $N_n/N_p\simeq 0.54$ at the center of the Sun~\cite{Serenelli:2009yc}, we can translate the LMA-Dark $2 \sigma$ band found in Refs.~\cite{Esteban:2018ppq, Gonzalez-Garcia:2013usa} into
\begin{equation} 
    2 \lesssim \varepsilon_{\mu \mu}^\text{medium}=\varepsilon_{\tau \tau}^\text{medium} \lesssim 3\,,
\end{equation} 
which translates into
\begin{equation} 
    g_{Z'}=(6.5-8.0)\times 10^{-5} \frac{m_{Z'}}{10~{\rm MeV}} \left(-\frac{1}{c}\right)^{1/2}.
\end{equation} 
Of course, there is also the standard LMA solution with $\theta_{12}<\pi/4$ that requires~\cite{Esteban:2018ppq, Gonzalez-Garcia:2013usa}
\begin{equation} 
    -0.081<\varepsilon_{\mu\mu}^\text{medium}=\varepsilon_{\tau\tau}^\text{medium}<1.422\,,
\end{equation} 
which, for $N_n/N_p \simeq 0.54$, corresponds to
\begin{equation} 
    -3\times 10^{-9}\left(\frac{m_{Z'}}{10~{\rm MeV}}\right)^2 < c\, g_{Z'}^2 < 1.7 \times 10^{-10} \left(\frac{m_{Z'}}{10~{\rm MeV}}\right)^2.
\end{equation} 

In our model, the coupling to the muon is given by $g_\mu\equiv g_{Z'}-e\, \epsilon$ which can be rewritten as 
\begin{equation}
    g_\mu = g_{Z'}\left[1-c\left(1-\frac{1}{\tan \eta}\right)\right] = \left[\frac{2\sqrt{2}\, G_F\, m_{Z'}^2}{-c}\, \frac{N_e}{N_n+N_p}\, \varepsilon_{\mu\mu}^\text{medium}\right]^{1/2} \left[1-c \left(1-\frac{1}{\tan \eta}\right)\right].
\end{equation}
In the limit $|c|\ll 1$, $g_\mu\simeq g_{Z'}$ as expected.
To explain the $(g-2)_\mu$ anomaly, $g_\mu$ should be in the range found in Ref.~\cite{Amaral:2021rzw}. For example, if $m_{Z'}\sim 10$~MeV, the $2\sigma$ band compatible with $(g-2)_\mu$ is
\begin{equation}
    g_\mu = (3.5-7)\times 10^{-4}\,. \label{gMU}
\end{equation} 
In the next section, we discuss the various bounds on the model and find the parameter range that can lead to interesting phenomenology.

If $\epsilon$ does not vanish, $Z'$ can be produced by its coupling to electrons. Furthermore, if $Z'$ is lighter than $2\, m_\mu$, the main $Z'$ decay modes are into  $\nu_\mu\, \bar{\nu}_\mu$,  $\nu_\tau\, \bar{\nu}_\tau$ and $e^-\, e^+$. Up to corrections of order of $(m_e/m_{Z'})^2$
\begin{equation} 
    \text{Br}(Z'\to e^-e^+)=\frac{(e\, \epsilon)^2}{(e\, \epsilon)^2+g_{Z'}^2}\,,   \  \  \ {\rm and} 
\  \  \   \text{Br}(Z'\to {\rm invisible})=\frac{g_{Z'}^2}{(e\, \epsilon)^2+g_{Z'}^2} \  .
\end{equation} 
As discussed in the next section, the NA48/4 experiment strongly constrains the scenario in which $Z'$ can be produced by the $\pi^0$ decay with subsequent decay into a pair $e^-e^+$. On the other hand, the NA64 experiment constraints $Z'$ that can be produced by its coupling to electrons and then decay invisibly. Motivated by saving the dark-photon solution to the $(g-2)_\mu$ anomaly, Ref.~\cite{Mohlabeng:2019vrz} suggests opening up a semivisible decay mode for $Z'$ to avoid these bounds. 

In the following, we suggest an alternative detour to these bounds by augmenting the model such that $Z'$ predominantly decays into a pair of intermediate scalars $\varphi\, \bar{\varphi}$ that, in turn, decay to pairs $e^-\, e^+$. We will see that this mechanism also gives mass to the $Z'$ boson. For $\varphi$ lighter than 10~MeV decaying to a pair $e^-\, e^+$, there are strong bounds from E774 and E141~\cite{Bross:1989mp} so  we take $\varphi$ to be heavier than 10~MeV. As a result,  the mass of $Z'$ should be larger than 20~MeV.
We assign a $U_\text{NEW}(1)$ charge $c_\varphi \gg 1$ to the $\varphi$ scalars, obtaining the coupling
\begin{equation} 
    c_\varphi\, g_{Z'}\, Z'_\mu \left[i(\varphi^*\, \partial^\mu \varphi )+\text{H.c.}\right] ,
\end{equation} 
which leads to a partial decay width
\begin{equation} 
    \Gamma\left(Z'\to \varphi\, \bar{\varphi}\right)=\frac{c_\varphi^2\, g_{Z'}^2\, m_{Z'}}{48\pi} \left(1 - 4\,\frac{m_\varphi^2}{m_{Z'}^2}\right)^{3/2}.
\end{equation} 
It is important to note that, despite $c_\varphi$ being large, we are interested in a range of parameters where the coupling of $\varphi$ to $Z'$, $c_\varphi\, g_{Z'}$ is  very small, and well within the perturbative range.
In the limit $c_\varphi^2\, g_{Z'}^2 \gg e^2 \epsilon^2$, the branching ratios can be rewritten as
\begin{equation} 
    \text{Br}(Z'\to e^-e^+)= \frac{8\, (e\, \epsilon)^2}{c_\varphi^2g_{Z'}^2(1-4m_\varphi^2/m_{Z'}^2)^{3/2}}   \  \  \  {\rm and} \  \  \   \text{Br}(Z'\to {\rm invisible})= \frac{8}{c_\varphi^2(1-4m_\varphi^2/m_{Z'}^2)^{3/2}}.
\end{equation} 

As we shall see in the next section, to avoid the NA64 bounds, a $Z'$ with the energy of $\sim 30$~GeV should be able to decay to $\varphi\, \bar{\varphi}$ before traveling a distance of $\sim 1$~cm. Similarly, the $\varphi$ produced should decay before traveling more than $\sim 1$~cm.  That is,
\begin{equation} 
    \tau_{Z'} < 3\times 10^{-14}\, \frac{m_{Z'}}{30~{\rm MeV}}\, \frac{30~{\rm GeV}}{E_{Z'}}~{\rm sec} \ \ \ \ {\rm and}  \ \ \ \ \  \tau_{\varphi} < 3\times 10^{-14}\, \frac{m_{\varphi}}{30~{\rm MeV}}\, \frac{30~{\rm GeV}}{E_{\varphi}}~{\rm sec}\,,
\end{equation} 
and hence,
\begin{equation}\label{lowerCPHI}
    c_\varphi\, g_{Z'}> 3.3\times 10^{-4}\, \frac{30~{\rm MeV}}{m_{Z'}}\, \left(1 - 4\,\frac{m_\varphi^2}{m_{Z'}^2}\right)^{-3/4}.
\end{equation}  
For $\varphi$ to decay into $e^- e^+$, it should be coupled to electrons. A direct coupling would break both the $U_\text{NEW}(1)$ gauge symmetry and the electroweak symmetry. Therefore, we introduce a second $\varphi'$ with the same charge as $\varphi$ and heavier than $Z'$. Furthermore, we add a new inert Higgs doublet $\Phi$ with a large coupling to $e^-e^+$ via the terms
\begin{equation} 
    \lambda\, \varphi^\dagger\, \varphi'\, H\cdot \Phi + \lambda_e\, \mathbf{\bar{e}_R}\, \Phi^\dagger\, \mathbf{L_e} \  ,
\end{equation} in which $ \mathbf{L_e}=(\nu_e \,\, \mathbf{e_L} ) $.
Since the SM Higgs coupling to electrons is very suppressed, we need this new $\Phi$ with a relatively large Yukawa coupling $\lambda_e$ to ensure a fast decay of $\varphi \to e^-e^+$ with $\tau_\varphi\sim 10^{-14}$ sec. With such a short lifetime, the bounds from E177 can also be relaxed~\cite{Bjorken:1988as} because decays occur before $\varphi$ or $Z'$ reach the detector.   The vacuum expectation value of $\varphi'$ breaks the $U_{NEW}(1)$ symmetry and gives mass to the $Z'$ boson
\begin{equation} 
    m_{Z'}=c_\varphi\, g_{Z'}\, \langle \varphi'\rangle \  .
\end{equation} 
Furthermore, along with $\langle H\rangle$, it leads to the mixing of $\varphi$ with the neutral component of $\Phi$ given by
\begin{equation} 
    \sin \beta=\frac{\lambda\, \langle H\rangle\, \langle \varphi'\rangle}{m_{\Phi^0}^2-m_\varphi^2}\,,
\end{equation} 
and, therefore, to an effective coupling of the form
\begin{equation} 
    \lambda_{\varphi e}\, \varphi^\dagger\, \mathbf{\bar{e}\, e}  \ \ \ \ {\rm with } \ \ \ \ \lambda_{\varphi e}\equiv \lambda_e\, \sin\beta \ .
\end{equation} 
For $\tau_\varphi \sim 10^{-14}$~sec, $\lambda_{\varphi e}$ should be of the order of $4\times 10^{-4}\, (10~{\rm MeV}/m_\varphi)^{1/2}$. Furthermore, $\Phi^0$ should be heavier than $\sim 400$~GeV to avoid present bounds from direct searches at colliders. Taking $\lambda_e \sim 0.1$, $\sin \beta$ should be of the order of $10^{-3}$. Note that such mixing is small enough not to cause an unnaturally large contribution to the $\varphi$ mass: $m_\varphi \gg m_\Phi\, \sin^2 \beta$. We can then write
\begin{equation} 
    \lambda =0.026\ \frac{\sin\beta}{10^{-3}}\, \frac{30~{\rm MeV}}{m_{Z'}}\, \frac{c_\varphi g_{Z'}}{8.5\times 10^{-4}}\, \frac{m_{\Phi^0}^2}{(400~{\rm GeV})^2}
	\  . \label{lambda}
\end{equation} 
Finally, to allow $\lambda$ to remain in the perturbative range, it is necessary that $c_\varphi\, g_{Z'}<{\rm few}\times 10^{-2}$.

 \section{The bounds \label{bound}}
For the values of the $g_{Z'}$ coupling of interest for NSI or for $(g-2)_\mu$, $Z'$ reaches thermal equilibrium in the early universe with the plasma. If $Z'$ is lighter than $\sim 5$~MeV, $Z'$ and/or its decay products can contribute significantly to the extra relativistic degrees of freedom in which there are strong bounds from CMB and BBN~\cite{Fields:2019pfx}. Therefore, we focus on the case where $m_{Z'} > 5$~MeV.
Now, we present a compilation of the most stringent bounds relevant to the present scenario.

{\it Bounds from beam dump experiments, meson decays, and scattering experiments:} \\
In the absence of $\varphi$, two regimes can be distinguished for $m_{Z'}\sim 10$~MeV.
\begin{enumerate} 
    \item $g_{Z'}^2\gg e^2\, \epsilon^2$:  In this case, $Z'$ decays mainly into $\nu_\mu\, \bar{\nu}_\mu$ and $\nu_\tau\, \bar{\nu}_\tau$. Thus, $Z'$ would appear as missing energy in experiments such as BaBaR~\cite{BaBar:2017tiz} and NA64~\cite{Banerjee:2019pds, Andreev:2022hxz}, where $Z'$ can be produced by its coupling to electrons (for example, by $e^-\,e^+\to \gamma\, Z'$ or electron bremsstrahlung). These experiments established an upper bound $\epsilon \lesssim \mathcal{O}\left(10^{-5}\right)$ for $m_{Z'}\sim 10$~MeV. As invisible decay modes dominate over visible decay modes, the bounds of beam dump experiments on $g_{Z'}$ and/or on $\epsilon$ are relaxed.
    The $Z'$ coupling to neutrinos can appear in meson decays such as $K^+\to \mu^+\, \nu\, Z'$. Using the constraint of E949 on $K^+\to \mu^+ + {\rm missing~energy}$~\cite{E949:2016gnh}, an upper bound on the coupling of $Z'$ to $\nu_\mu$ can be extracted~\cite{Bakhti:2017jhm}. With an improved constraint from NA62 on such decay modes~\cite{NA62:2021bji}, the bound for the mass range $m_{Z'}^2/m_K^2$ can be rewritten as 
    \begin{equation}\label{Kaon}
        g_{Z'}<0.003 \left(\frac{m_{Z'}}{5~{\rm MeV}}\right).
    \end{equation} 
    Moreover, from the bound on $\pi^0 \to Z'\, \gamma$, an upper bound of $\mathcal{O}\left(10^{-3}\right)$ on the coupling of $Z'$ to quarks is obtained~\cite{NOMAD:1998pxi, Farzan:2016wym}.
    
    \item    $g_{Z'}^2\ll e^2\, \epsilon^2$: In this case, a $Z'$ with mass $m_{Z'} \sim \mathcal{O}\left(10\right)$~MeV decays mainly into pairs $e^-\, e^+$, relaxing the bound from NA64.  Instead, the bounds from beam dump experiments apply.
    For $m_{Z'}\sim 10$~MeV, the strongest upper bound on $\epsilon$  comes from the NA48/2 experiment~\cite{NA482:2015wmo}. The bound on $\epsilon$ versus $m_{Z'}$ fluctuates violently between $5 \times 10^{-4}$ and $10^{-3}$.
    For the time being, the parameter space where the $(g-2)_\mu$ anomaly can be fitted (that is, $g_{Z'} \sim 7 \times 10^{-4}$) is experimentally allowed. Interestingly, such parameter space will be probed by future experiments such as MESA~\cite{Beranek:2013yqa}, VEPP-3~\cite{Wojtsekhowski:2009vz, Wojtsekhowski:2012zq}, and DARKLIGHT~\cite{Freytsis:2009bh}.
\end{enumerate}

Opening the decay mode $Z' \to \varphi\, \bar{\varphi}$ described at the end of the previous section, the bounds from NA64 and NA48/2 can be relaxed. In NA64, an electron beam of $100 \pm 3$~GeV~\cite{Banerjee:2019pds} is sent to an ECAL target. If the energy deposited within a few radiation lengths is less than 50~GeV, the signal is interpreted as $e^- +{\rm nucleus} \to e^-\, Z'\, X$, with $Z'\to {\rm missing~energy}$.
In our model, $Z' \to \varphi\, \bar{\varphi}$ and $\varphi \to e^-\, e^+$ within a few centimeters, so the entire energy of the initial $e^-$ entering the target at NA64 will be deposited at the ECAL within a few radiation lengths, so the NA64 bound will be relaxed. In NA48/2, the signal is $e^-\, e^+\, \gamma$ from $\pi^0$ decays, and events in which the invariant mass of the three final tracks significantly deviates from $m_{\pi^0}$ are vetoed.
Thus, $e^-\, e^+$ from the $\varphi$ decay will be vetoed. To recast the bound from BaBaR~\cite{BaBar:2017tiz} and NA64~\cite{Andreev:2022hxz} on the coupling of $Z'$ to the electron, we should take into account the expression of $\text{Br}(Z'\to {\rm invisible})$ in the present model. Similarly, the bound from NA48/2~\cite{NA482:2015wmo} should be recasted by considering $\text{Br}(Z'\to e^-\, e^+)$ in this model.
In the simple kinetic mixing model (that is, when $g_{Z'} = 0$), the $\pi^0$ decay rate to a photon and a dark photon is proportional to $e^2\, \epsilon^2\, (q_u^2 + q_d^2)^2 = (5/9)^2\, e^2\, \epsilon^2$. In our model where both $g_{Z'}$ and $\epsilon$ are nonzero, it will be given by $[e\, q_u\, (e\, \epsilon\, q_u + c\, g_{Z'}/3) + e\, q_d\, (e\, \epsilon\, q_d + c\, g_{Z'}/3)]^2$. Thus, the branching ratio of a $\pi^0$ decaying to a photon and a $Z'$ will be given by the same formula for pure kinetic mixing, replacing $\epsilon^2$ with $(\epsilon - c\, g_{Z'}/(5\, e))^2$. Furthermore, in our model, $\text{Br}(Z' \to e^-\, e^+)$ is not 1 so the bound on the square of mixing found by NA48/2 should be interpreted as a limit on $[\epsilon -c\, g_{Z'}/(5\, e)]^2 \times {\rm Br}(Z' \to e^-\, e^+)$.
In the $B-L$ model considered in Ref.~\cite{Andreev:2022hxz}, ${\rm Br}(Z' \to {\rm invisible})|_{B-L}={\rm Br}(Z' \to \nu \bar{\nu})/[{\rm Br}(Z' \to e^-e^+)+{\rm Br}(Z' \to \nu \bar{\nu})]=3/5$. In NA64, $Z'$ is produced by its coupling to electrons, which for us is $e\, \epsilon$. As a result, the upper bound on the square of the $B-L$ coupling found in Ref.~\cite{Andreev:2022hxz} should be interpreted as an upper bound on $(e\, \epsilon)^2 \times {\rm Br}(Z' \to {\rm invisible})|_\text{ours} / {\rm Br}(Z' \to {\rm invisible})|_{B-L}$.

A dedicated search using experiments such as BaBaR, NA64, or NA48/2 may be able to test our model where $Z'$ production leads to the emission of two pairs (rather than one pair) of $e^-\, e^+$. As mentioned above, the bounds from the E177, E774, and E141 beam dump experiments can be avoided in our model.
Finally, we note that $Z'$ bosons can also be probed at the intensity and lifetime frontier experiments such as FASER, FASER2, DUNE, and the ILC~\cite{Asai:2022zxw}.\\

{\it Coherent Elastic Neutrino-Nucleus Scattering (CE$\nu$NS) experiments:} \\
In our model, since the coupling of $\nu_e$ to $Z'$ is zero (i.e., $\varepsilon_{ee}=0$), the reactor CE$\nu$NS experiments such as Dresden~II~\cite{Denton:2022nol} or CONUS do {\it not} constrain the model. 
However, we expect bounds from CE$\nu$Ns experiments with a muon decay source such as COHERENT, as well as from direct dark matter search experiments sensitive to solar neutrinos.
The cross section of the CE$\nu$NS process $\nu_\mu +{\rm nucleus} \to \nu_\mu
+{\rm nucleus}$ is proportional to
\begin{equation} 
    Q_\mu^2 = Z\left(g_p^V + \varepsilon_{\mu\mu}^p\, \frac{m_{Z'}^2}{m_{Z'}^2-t}\right) + N \left(g_n^V + \varepsilon_{\mu\mu}^n\, \frac{m_{Z'}^2}{m_{Z'}^2-t}\right),
\end{equation} 
with $t$  being a Mandelstam variable.  $g_p^V=1/2-2\sin^2\theta_W$ and $g_n^V=-1/2$ are the vector couplings of the standard $Z$ gauge boson to protons and neutrons, respectively. Furthermore, $Z$ and $N$ are the numbers of protons and neutrons in the target nucleus. With $\varepsilon_{\mu\mu}^n=-(Z/N)\, \varepsilon_{\mu\mu}^p$ (that is, $\tan\eta = -Z/N$), the NSI effect completely cancels out. For CsI and argon, these ratios are $\tan \eta=-0.7$ and $\tan \eta=-0.8$, respectively.
The results of Ref.~\cite{Chaves:2021pey} confirm this argument.
For our model, the allowed range of NSI can be even larger due to the suppression $|m_{Z'}^2/(m_{Z'}^2-t)|< 1 $. 
In our figures, we take the average for argon and CsI: $\tan \eta=-0.75$.
For other target materials, this cancelation occurs at different values of $\tan\eta$.  For example, for silicon $\tan \eta=-Z/N=-1$. As a result, the change of the target material to silicon has the potential to test this degeneracy~\cite{Baxter:2019mcx}.

Note that there is also degeneracy under $Q_\mu\to -Q_\mu$. For $m_{Z'}^2 \gg t$, the latter transformation can take place for $Z\, \varepsilon_{\mu\mu}^p + N\, \varepsilon_{\mu\mu}^n = -2\, (Z\, g_p^V + N\, g_n^V)$. As a result, Ref.~\cite{Chaves:2021pey} finds a four-fold degeneracy. However, for $ m_{Z'}^2\sim |t|$, the $Q_\mu \to -Q_\mu$ degeneracy (but not the $\varepsilon^n/\varepsilon^p=-Z/N$ degeneracy) can be solved in principle by studying the dependence of the recoil energy.\\

{\it Borexino results for the scattering of neutrinos off electrons:}\\
In our model, $\varepsilon_{\mu\mu}^e = \varepsilon_{\tau\tau}^e \ne 0$ so the bounds from the Borexino experiment in Ref.~\cite{Coloma:2022umy} have to be taken into account. 
Rewriting Eq.~\eqref{eee} as 
\begin{equation} 
    \varepsilon^e_{\mu\mu} = \varepsilon^e_{\tau\tau} = \varepsilon^\text{medium}_{\mu\mu}\left(1-\frac{1}{\tan \eta}\right) \frac{N_e}{N_n+N_p}\,,
\end{equation} 
it can be realized that the Borexino bound on $|\varepsilon^e_{\mu\mu}= \varepsilon^e_{\tau\tau}|<2$~\cite{Coloma:2022umy} implies that the LMA-Dark solution at $\tan \eta=-0.8$ to $-0.7$ is excluded regardless of the values of $m_{Z'}$, $c$ and other parameters. Within our model, LMA-Dark can be compatible with the Borexino bound only for $\tan \eta > 0.5$ or for $\tan \eta<-37$. 
However, a large NSI with $\varepsilon^e_{\mu\mu}= \varepsilon^e_{\tau\tau}\sim 1$ still escapes the Borexino bound even at $\tan \eta \sim -0.75$. Such a large NSI will induce a significant deviation from the standard MSW prediction for the low-energy part of the $^8$B solar neutrino spectrum, despite the vanishing contribution to CE$\nu$NS.\\

{\it White dwarf cooling:}\\
Large effective couplings between electrons and neutrinos could lead to rapid cooling of white dwarfs~\cite{Dreiner:2013tja}. 
As shown in Ref.~\cite{Bauer:2018onh}, white dwarf cooling sets a bound
\begin{equation} 
    \frac{2\, g_{Z'}^2\, \epsilon}{3\, m_{Z'}^2} < 1.12\times 10^{-5}~{\rm GeV}^{-2},
\end{equation} 
which is considerably weaker than the other relevant bounds discussed above.\\

{\it Self-interaction of neutrinos in supernovae:}\\
The $g_{Z'}$ coupling could lead to resonant annihilation processes $\nu_\mu\bar{\nu}_\mu,\, \nu_\tau\bar{\nu}_\tau \to Z' \to \nu_\mu\bar{\nu}_\mu,\, \nu_\tau\bar{\nu}_\tau$, such that at $m_{Z'}\sim 30$~MeV, even for $g_{Z'} \sim 10^{-5}$, the mean free path of neutrinos (antineutrinos) will be shorter than that of SM scattering off nucleons~\cite{Cerdeno:2023kqo}. This consideration has been invoked in Ref.~\cite{Kamada:2015era} to evaluate the duration of the burst using the simplified formula $\Delta t \sim R_\text{core}^2/({\rm mean ~ free ~ path})$ and to set a bound on the coupling $g_{Z'}$ from the measured duration of the SN1987a neutrino burst. However, as shown in Ref.~\cite{Dicus:1988jh}, when neutrinos are isotropically distributed, self-interactions cannot prolong the duration of neutrino bursts~\cite{Cerdeno:2023kqo}.\\

{\it $B$ physics:}\\
As mentioned above, since in our model the $U_\text{NEW}(1)$ charges of the third generation of quarks are different from those of the first two generations, in the mass basis, the quarks obtain a Flavor-Changing Neutral Current (FCNC) coupling to $Z'$.  After integrating out the $Z'$ boson, we obtain an effective coupling of the form
\begin{equation} 
    H_\text{eff}=\frac{g_{Z'}^2\, \pi}{2\, m_{Z'}^2}\, V_{ti}\, V_{tj}^*\, \frac{(3c - 2)}{3} \left(\bar{d}_i\, \gamma^\mu\, P_L\, d_j\right) \left(\bar{l}\, \gamma^\mu\, l\right), \label{FCNC}
\end{equation} 
where $l\in \{\mu, \tau\}$, $d_i$, $d_j \in \{d, s, b\}$, and $V_{ti}$ and $V_{tj}$ are the elements of the third row of the CKM matrix. Note that although we start with a non-chiral coupling of $Z'$ to fermions, the FCNC coupling in Eq.~\eqref{FCNC} is chiral because it originates from the quark-mass term, which mixes chiralities. That is, we have the freedom to choose a basis where the right-handed quark couplings to $Z'$ remain diagonal and attribute all FCNC to left-handed down quarks. Here, we use the common notation used in the literature of $b$ anomalies~\cite{Crivellin:2015lwa}, 
\begin{equation}\label{C9}
    C_9 =-\frac{g_{Z'}^2\, \pi}{\sqrt{2}\, m_{Z'}^2}\, \frac{1}{\alpha_\text{EM}\, G_F}\frac{3\, c-2}{3}\,.
\end{equation} 
In the case where $C_9 \sim -1$, the so-called $b$-anomalies can be explained~\cite{b-anomaly}. However, very recent LHCb results seem to be compatible with the SM, reducing the need for new physics~\cite{LHCb:2022qnv, LHCb:2022zom}.
We should note that in our model $m_{Z'}\ll m_b$, and therefore the effective action formalism cannot be used to calculate $b\to s\, \mu^+\, \mu^-$. In fact, the contribution of our model to the amplitude of this process will be suppressed by a factor of $m_{Z'}^2/(m_{Z'}^2-q^2)$ relative to $C_9$, where $q^2$ is the invariant mass of the final muon pair. In our model,
\begin{equation} 
    C_9 \times \frac{m_{Z'}^2}{m_{Z'}^2-q^2}=- 2 \left(\frac{g_{Z'}}{3.5 \times 10^{-4}}\right)^2\frac{{\rm GeV}^2}{q^2}\, \frac{3\, c-2}{-2}\,.
\end{equation} 
Note that for $g_{Z'}$ in the range that explains $(g-2)_\mu$, the deviation in the low-energy bins of $q^2$ can be significant.  Taking $c=2/3$, the $U_\text{NEW}(1)$ charges of the quarks of all generations will be equal, so $b\to s\, \mu^+ \mu^-$ cancels out. The anomaly cancelation can also be fulfilled by adding more generations of fermions charged under $U_\text{NEW}(1)$.  
Notice that in our model, the FCNC contribution to $b \to d$ is suppressed by one more order of magnitude, that is, by $V_{td}/V_{ts}$.  

\begin{figure}[t!]
    \def\sepf{0.64}
	\centering
    \includegraphics[scale=\sepf]{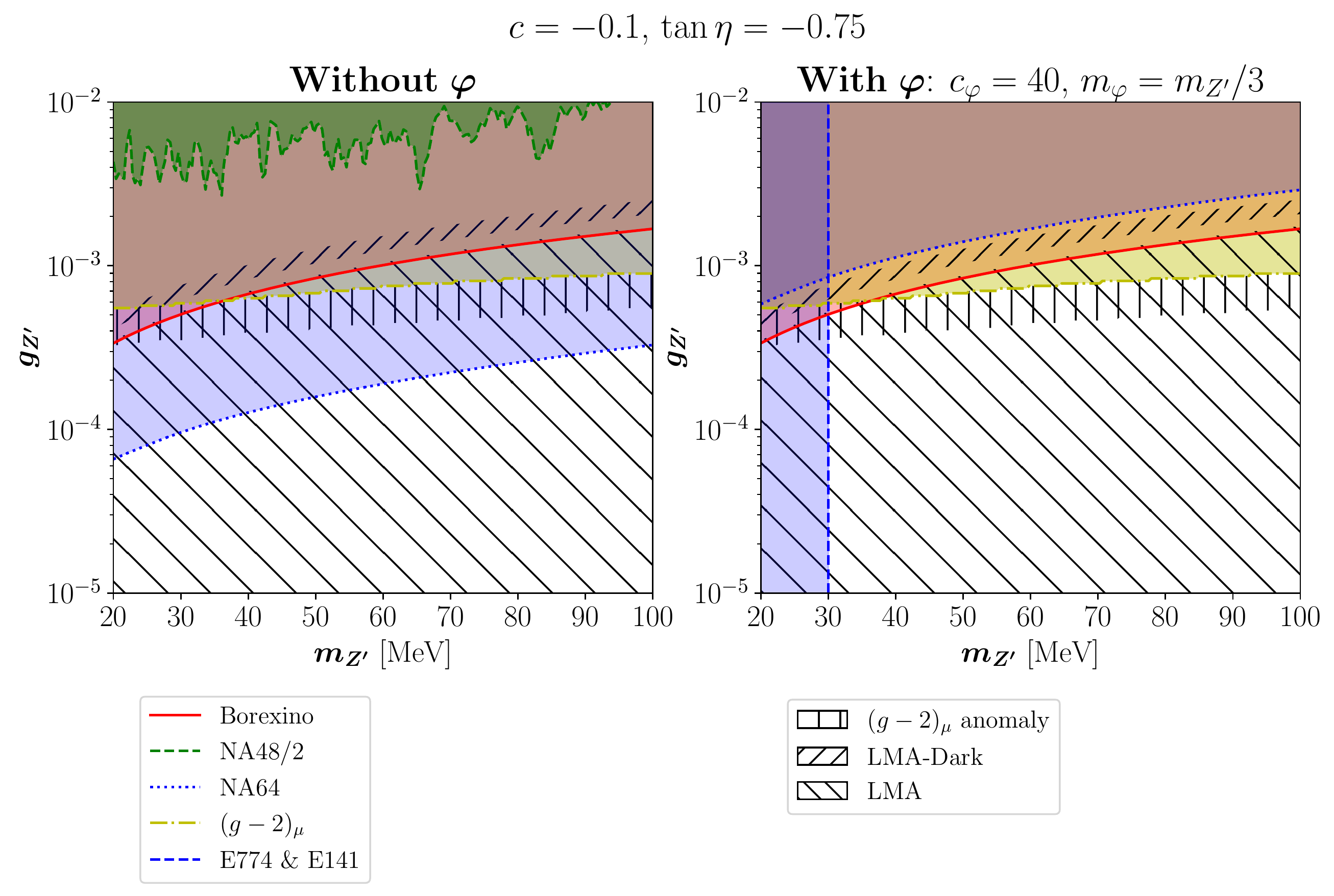}
    \caption{Bounds on the parameter space of the model for $c = -0.1$ and $\tan\eta = -0.75$.   The colored regions are excluded by various experiments as indicated in the legend and described in the text. The vertically dashed regions are favored by the $(g-2)_\mu$ anomaly. The LMA-Dark solution to the solar neutrino anomaly, as well as the bounds on NSI the neutrino oscillation data~\cite{Gonzalez-Garcia:2013usa, Esteban:2018ppq}, are indicated by diagonally dashed lines. The right and left panels correspond to the variations of the model with and without $\varphi$, respectively (see Sect.~\ref{model} for a description). In the right panel, we have  taken $c_\varphi = 40$ and $m_\varphi = m_{Z'}/3$.  With this ratio, for $m_{Z'}<30$~MeV ({\it i.e.,} to the left of the vertical line in the right panel), $\varphi$ would be  too light to avoid the bounds from E774 and E141~\cite{Bross:1989mp}.}
	\label{fig:Z_c_-0.1}
\end{figure} 

\begin{figure}[t!]
    \def\sepf{0.64}
	\centering
    \includegraphics[scale=\sepf]{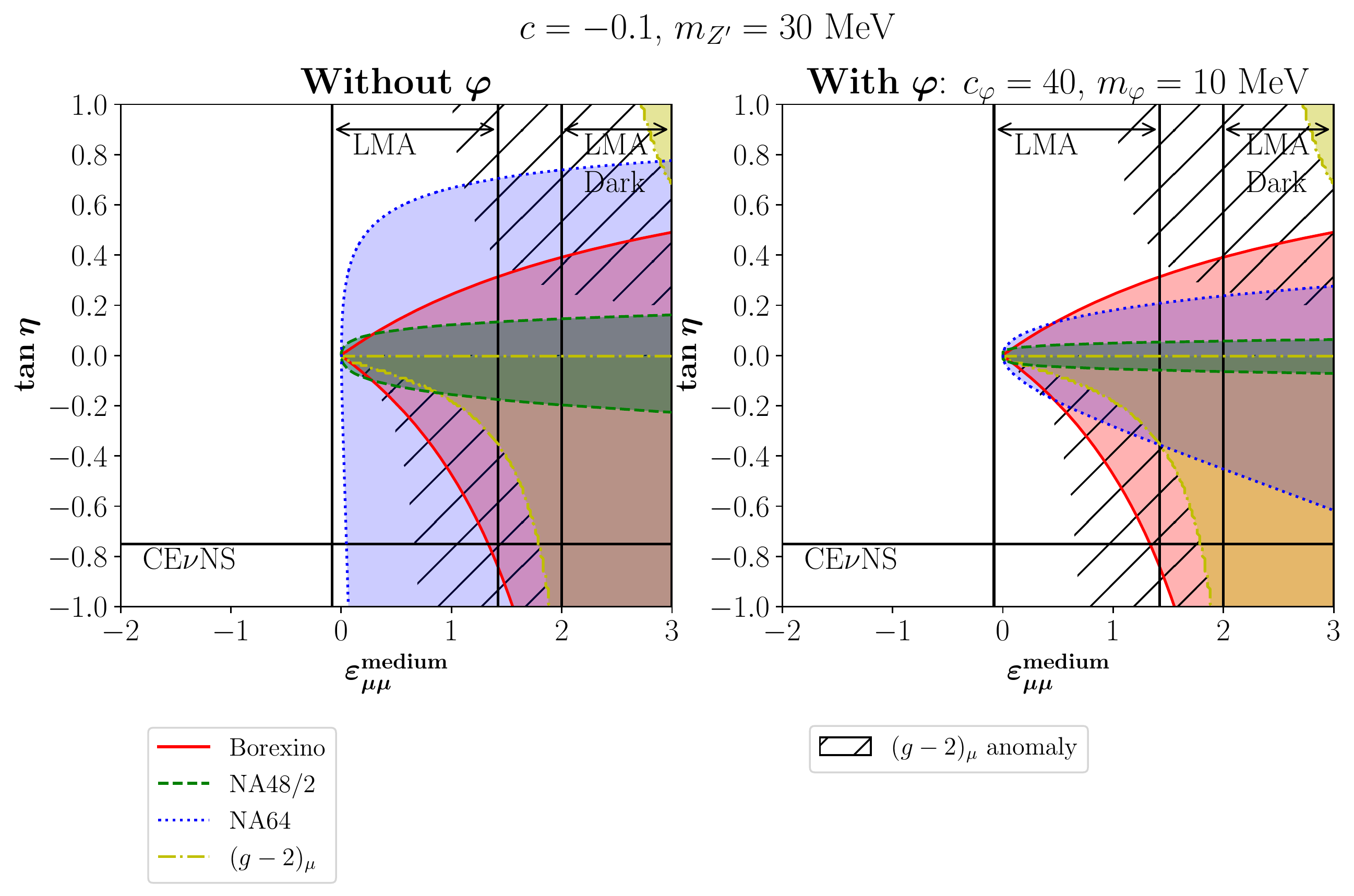}
    \caption{The same as Fig.~\ref{fig:Z_c_-0.1}, but projected in the $[\varepsilon_{\mu\mu}^\text{medium},\, \tan\eta]$ plane.
    The vertical bands correspond to the LMA and LMA-Dark solutions. The horizontal line depicts $\tan\eta = -0.75$ for which the contribution from the new physics to CE$\nu$NS is suppressed.}
	\label{fig:c_-0.1}
\end{figure}

\begin{figure}[t!]
	\def\sepf{0.64}
	\centering
	\includegraphics[scale=\sepf]{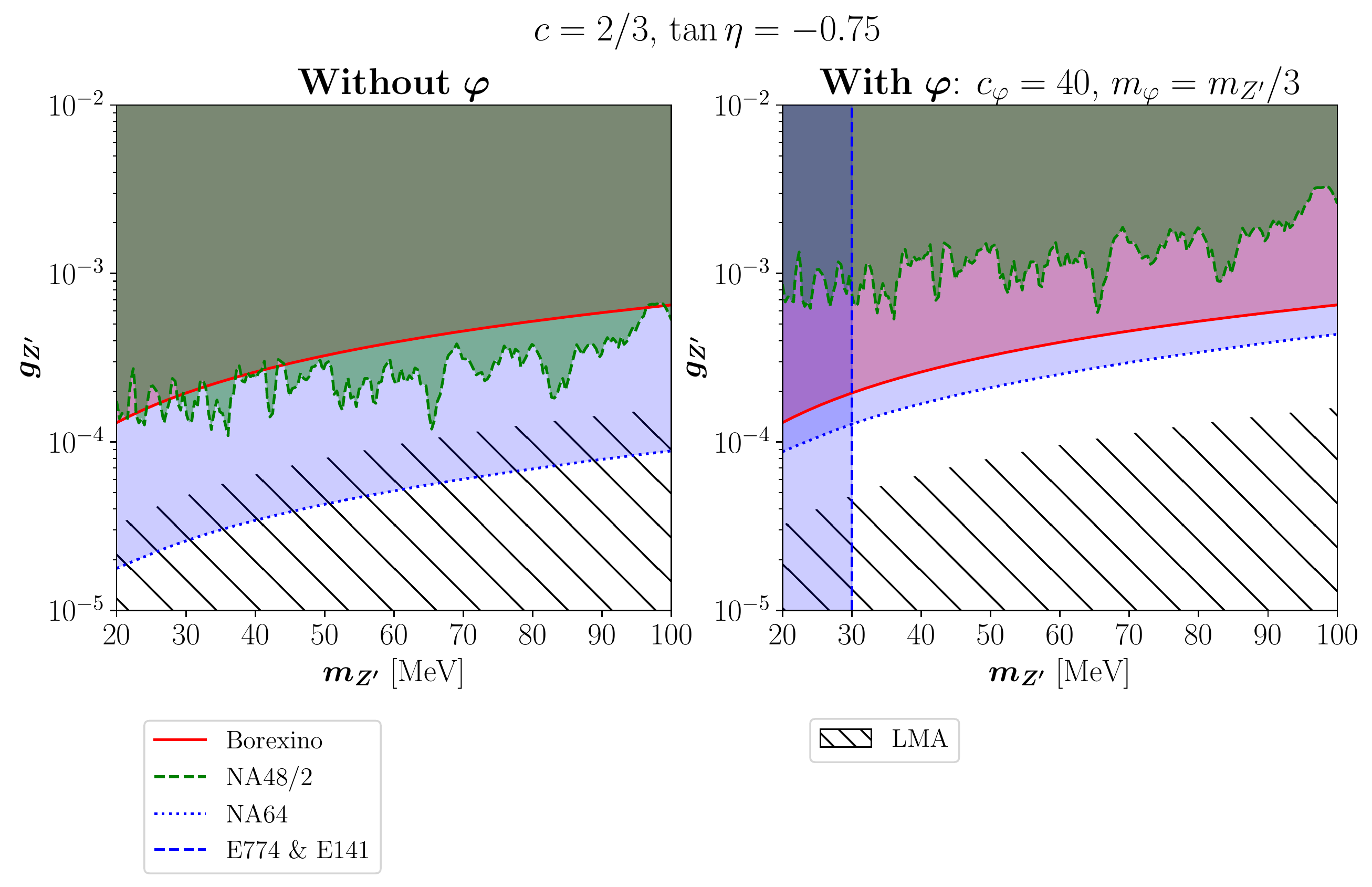}
	\caption{The same as Fig.~\ref{fig:Z_c_-0.1}, but for $c = 2/3$.}
	\label{fig:Z_c_2o3}
\end{figure}

\begin{figure}[t!]
	\def\sepf{0.64}
	\centering
	\includegraphics[scale=\sepf]{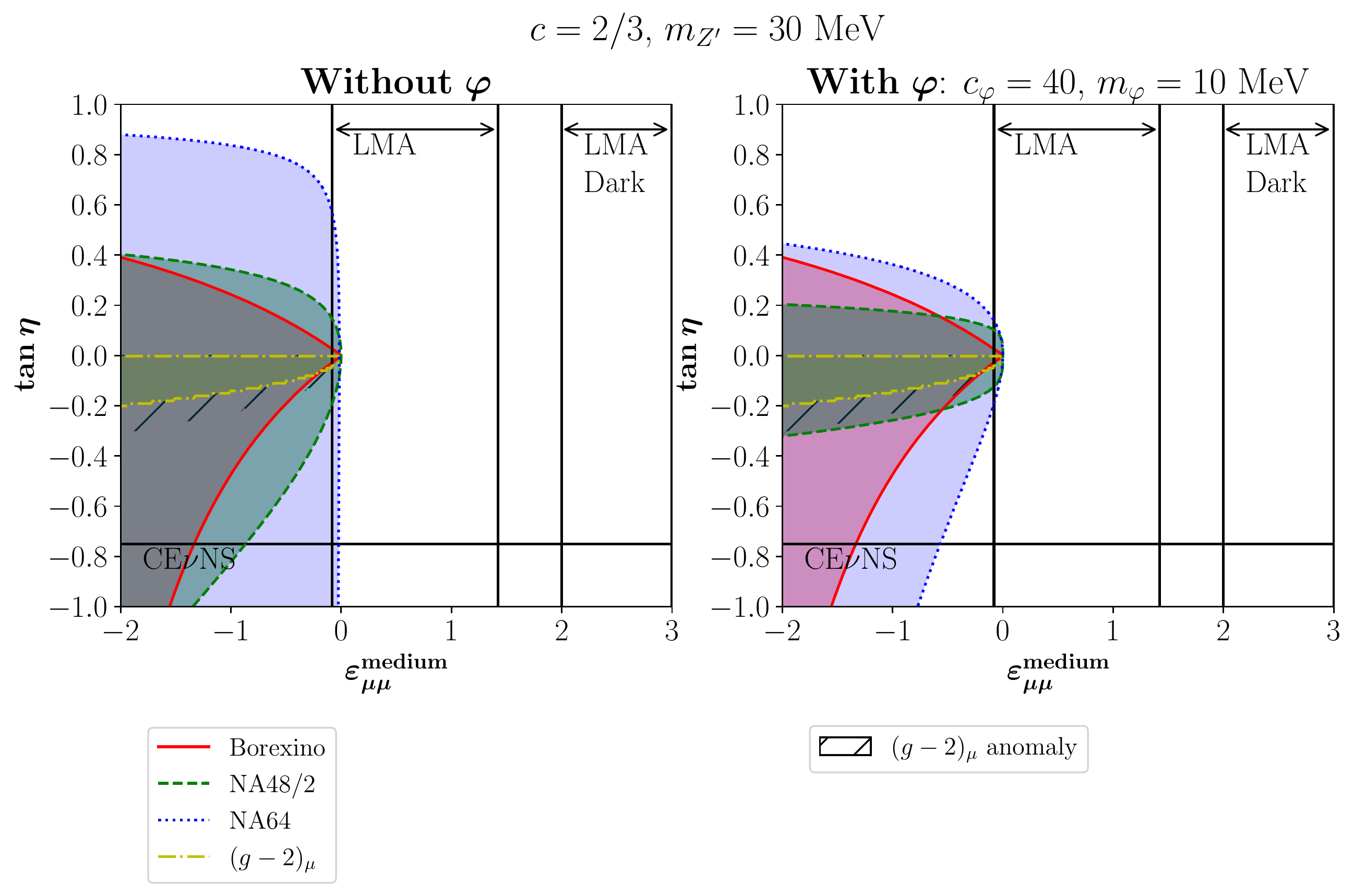}
	\caption{The same as Fig.~\ref{fig:c_-0.1}, but for $c = 2/3$.}
	\label{fig:c_2/3}
\end{figure}

Figures~\ref{fig:Z_c_-0.1} to~\ref{fig:c_2/3} summarize all the relevant bounds discussed above.
Figure~\ref{fig:Z_c_-0.1} shows the bounds for $c = -0.1$ and $\tan\eta = -0.75$ in the $[m_{Z'},\, g_{Z'}]$ plane. The value of $\tan \eta=-0.75$ is chosen because at this value the contribution of new physics to CE$\nu$NS cancels out. The colored regions show the excluded parameter ranges as follows: Borexino measurements of solar neutrino scattering off electrons (solid red), searches for $Z'$ decaying into $e^+e^-$ at NA48/2 (dashed green), searches for invisible $Z'$ decays at NA64 (dotted blue), the upper bound $m_\varphi < 10$~MeV from the combination of E774 with E141 (vertical dashed blue), and the region where the contribution of new physics to $(g-2)_\mu$ exceeds the observed deviation from the SM prediction (dashed-dotted green).
In the vertical dashed area, our model provides an explanation for the $(g-2)_\mu$ anomaly.
Furthermore, the diagonally dashed regions correspond to the LMA and LMA-Dark solutions, for which $-0.081 < \varepsilon_{\mu\mu}^\text{medium} < 1.422$ and $2 < \varepsilon_{\mu\mu}^\text{medium} < 3$, respectively.
The right panel of Fig.~\ref{fig:Z_c_-0.1} shows the case with an additional scalar $\varphi$, assuming $c_\varphi = 40$ and $m_\varphi = m_{Z'}/3$. With this value of $c_\varphi$, the perturbativity limit discussed at the end of Sect.~\ref{model} is still satisfied.

Similar information projected in the $[\varepsilon_{\mu\mu}^\text{medium},\, \tan\eta]$  plane is presented in Fig.~\ref{fig:c_-0.1}.
The vertical bands correspond to the LMA and LMA-Dark solutions, and the horizontal line represents $\tan\eta = -0.75$, at which the contribution from new physics to CE$\nu$NS vanishes.
As discussed above, for $-37 \lesssim \tan \eta \lesssim 0.5$, the LMA-Dark solution cannot be compatible with the Borexino bound within our model, regardless of the values of the other parameters. 
However, for higher values of $\tan \eta$, we can have the LMA-Dark solution without conflict with other bounds. The CE$\nu$NS measurements will eventually test this solution with $\tan \eta>0.5$ even for light $Z'$~\cite{Denton:2018xmq, DeRomeri:2022twg}.  As shown in Fig.~\ref{fig:c_-0.1}, for $\tan \eta>0.5$, both the LMA-Dark and LMA bands have a significant overlap with the dashed area in which our model can explain the $(g-2)_\mu$ anomaly.
Fig.~\ref{fig:c_-0.1} also shows the conflict between LMA-Dark and the Borexino bound for negative values of $\tan \eta$. However, as seen in these figures for $c=-0.1$ and $m_{Z'}\sim 40$~MeV, values of $\varepsilon_{\mu\mu}^\text{medium}=\varepsilon_{\tau\tau}^\text{medium}\sim1$ can be compatible with the Borexino bound at $\tan\eta \sim -0.7$, with the values of $g_\mu$ also explaining the $(g-2)_\mu$ anomaly. Without $\varphi$, the NA64 results exclude this interesting part of the parameter range but can be revived by introducing $\varphi$, as demonstrated in the right panels of Figs.~\ref{fig:Z_c_-0.1} and~\ref{fig:Z_c_2o3}. This solution to the $(g-2)_\mu$ anomaly can be tested by $i)$ searching for $\varphi$ coupled to the electron in beam dump experiments; $ii)$ searching for the $\varepsilon_{\mu\mu}^\text{medium}=\varepsilon_{\tau\tau}^\text{medium}$  effects
in the spectrum of solar neutrinos especially around $E_\nu\sim 3$~MeV to be probed by the THEIA detector; $iii)$ searching for new physics in $b \to s\, \mu^+ \mu^-$ with a signature enhanced in lower bins of the $\mu^+ \mu^-$ invariant mass; and $iv)$, by a dedicated search for light $Z'$ producing two pairs of electron-positron.

From Eq.~\eqref{medium}, we observe that for $c<0$ ($c>0$), $\varepsilon_{\mu \mu}^\text{medium}=\varepsilon_{\tau \tau}^\text{medium}$ is positive (negative). The bound on the negative values of $\varepsilon_{\mu \mu}^\text{medium}=\varepsilon_{\tau \tau}^\text{medium}$ from the oscillation data is more stringent. For completeness, we have included Figs.~\ref{fig:Z_c_2o3} and~\ref{fig:c_2/3}  with a positive $c$:  $c = +2/3$. At this value of $c$, the quarks of the three generations have the same $U_{NEW}(1)$ charges, leading to a vanishing new contribution to FCNC and therefore to $b\to s\, \mu^+ \mu^-$.

\section{Conclusions and discussion\label{DIS}}
In the literature, there is a class of models based on flavor gauge symmetries with a MeV-ish gauge boson that lead to Non-Standard neutral current Interaction (NSI) between neutrinos and quarks. By gauging the baryon number, the couplings of the $u$ and $d$ quarks are equal since they share the same baryon number. For relatively light $Z'$, the contribution to CE$\nu$NS is suppressed, so the present CE$\nu$NS bounds allow relatively large NSI for $m_{Z'}<30$~MeV. However, these models can eventually be tested by improving the precision of the CE$\nu$NS experiments. As shown in Ref.~\cite{Chaves:2021pey}, to hide NSI from CE$\nu$NS, the ratio of $\tan \eta =\varepsilon^n/\varepsilon^p$ should have a certain value $\tan \eta=-Z/N\simeq -0.75$. In this paper, we have built a model that can produce NSI with arbitrary $\tan \eta$. The model is based on gauging a combination of the lepton and baryon numbers of different generations with a light gauge boson $Z'$ that mixes with the photon. The mixing breaks the equality of the couplings of the up- and down-quarks because they have unequal electric charges. Within this framework, the NSI couplings are lepton flavor-conserving. Since we do not gauge $L_e$, the NSI for $\nu_e$ and $\bar{\nu}_e$ (that is, $\varepsilon_{ee}$) remains zero, so the bounds from $\nu_e$ or $\bar{\nu}_e$ scattering (such as the ones from GEMMA~\cite{GEMMA}) can be evaded. However, because of the gauge boson mixing with the photon, non-standard interactions between the muon and tau neutrinos with the electron ({\it i.e.} $\varepsilon_{\mu\mu}^e$ and $\varepsilon_{\tau \tau}^e$, respectively) are unavoidable.  Thus, we expect an observable effect on the scattering of solar neutrinos off electrons at detectors such as Borexino. Within our model, the Borexino bound is not compatible with the LMA-Dark solution for $-37<\tan \eta<0.5$. However, we have found regions of the parameter space with $\tan \eta>0.5$ in which both LMA-Dark and a solution to the $(g-2)_\mu$ anomaly can be achieved. Interestingly, this parameter space range can be tested by CE$\nu$NS experiments exploiting spallation neutron sources.

We have focused on regions of the parameter space for which $\tan \eta\simeq -0.75$.  In this range, even large values for $\varepsilon^\text{medium}$ can be hidden from CE$\nu$NS experiments. We have found that $\varepsilon_{\mu\mu}^\text{medium}=\varepsilon_{\tau \tau}^\text{medium}\sim 1$ and a solution to $(g-2)_\mu$ can be obtained simultaneously.  If the invisible decay mode of $Z'$ dominates, the bound from NA64 rules out the $\tan \eta \sim -0.75$ range with large $\varepsilon_{\mu\mu}^\text{medium}\sim 1$. However, it becomes viable once the decay mode $Z'\to \varphi\, \bar{\varphi} \to e^-e^+e^-e^+$ is allowed.  The light $\varphi$ particles that decay into pairs $e^-e^+$ can be searched by beam dump experiments. Furthermore, $\varepsilon_{\mu\mu}^\text{medium} = \varepsilon_{\tau \tau}^\text{medium} \sim 1$ can be tested with future solar neutrino experiments. If in the solar neutrino data evidence for $\varepsilon_{\mu\mu}^\text{medium} = \varepsilon_{\tau \tau}^\text{medium} \sim 1$ is found without a corresponding signal at CE$\nu$NS, an interpretation would be $\tan \eta=-0.7$. Within our model, this also implies a distinct feature in the distribution of the invariant mass of the muon pair at $b \to s\, \mu^+ \mu^-$, which can be tested.

\subsection*{Acknowledgments}
NB received funding from the Spanish FEDER/MCIU-AEI under grant FPA2017-84543-P.
YF has received financial support from Saramadan under contract No.~ISEF/M/401439. She would like to acknowledge the support from the ICTP through the Associates Programme and from the Simons Foundation through grant number 284558FY19.
This project has received funding and support from the European Union's Horizon 2020 research and innovation programme under the Marie Sk{\l}odowska-Curie grant agreement No.~860881 (H2020-MSCA-ITN-2019 HIDDeN).


\end{document}